# THE DEFECT EFFECT ON THE ELECTRONIC CONDUCTANCE IN BINOMIALLY TAILORED QUANTUM WIRE

**H. S. Ashour[a], A. I. Ass'ad[b], M. M. Shabat[c]**

[a]Department of Physics, Al-Azhar University of Gaza, Gaza Strip, P.A.
E-mail: hashour@alazhar-gaza.edu
[b]Department of Physics, Al-Aqsa University, Gaza Strip, P.A.
E-mail: assad_assad@mail.alaqsa.edu
[c]Department of Physics, Islamic University of Gaza, Gaza Strip, P.A.
E-mail: shabat@mail.iugaza.edu.ps

## ABSTRACT

The paper considers the effect of the defects on the electronic transmission properties in binomially tailored waveguide quantum wires, in which each Dirac delta function potential strength have been weight on the binomial distribution law. We have assumed that a single free-electron channel is incident on the structure and the scattering of electrons is solely from the geometric nature of the problem. We have used the transfer matrix method to study the electron transmission. We found this novel structure has a good defect tolerance. We found the structure tolerate up to $\pm 20\%$ in strength defect and $\pm 5\%$ in position defect for the central Dirac delta function in the binomial distribution. Also, we found this structure can tolerate both defect up to $\pm 20\%$ in strength and $\pm 5\%$ in position dislocation.

## 1. INTRODUCTION

In the last decade, there was a growing interest in the electron conductance through one-dimensional scattering problems, especially in those cases where the potential is periodic structure with finite number of identical cells [1, 2]. Because of the remarkable advances in nano-technology and micro fabrication, it is possible to confine electrons in a conductor with a lateral extent of $100nm$ or less, resulting narrow quantum wire [3]. In these mesoscopic devices, the electron transport is best described by quantum mechanics. Miniature size of these devices eliminates the defect of scattering. At a low enough temperature, the motion of electrons through these devices is ballistic or quasiballistic and the electron-phonon interaction can be neglected. So that, the phase coherence length enlarges enough when compared with the device dimension. Mesoscopic devices can be considered as a coherent elastic scatterer [3]. Therefore, the electron transport properties, solely depends upon the geometrical structure of the quantum waveguide.

In recent years, there has been a growing interest in the electron transport through a sequence of Dirac delta function potential [4-9]. The researcher used different methods to study the electron transport in a waveguide quantum wire [10-12]. Recently, Ashour et al [13] has proposed a novel structure which is the binomially tailored waveguide quantum wires, in which each Dirac Delta function potential strength has been weight on the binomial distribution law. In this paper, we study the defects effect on the electronic conductance on the novel structure proposed by [13].

In section 2, we outline the transfer matrix method which connects the solutions at the ends of the waveguide quantum wires. We introduce the novel structure of the binomially tailored waveguide quantum wires. In section 3, we explore defect effects on the electronic conductance spectrum through the binomially tailored quantum wire. In this section, we have studied strength defect and dislocation defects on the central Dirac delta function in the distribution. Section 4 has been devoted to the conclusions of this study.

## 2. TRANSMISSION MATRIX THOUGH

### 2.1 Periodic Structure

In this context, we consider a finite periodic structure of Dirac delta function potential (Dirac Comb). Also, we assumed that the structure is narrow enough so that just single channel of electrons can be considered. In this treatment, we neglected electron-electron interaction, and we assume the temperature low enough so that electron-phonon interaction can be neglected as well. We assumed the scattering of electrons mainly form the geometrical structure of the potential. The potential can be written as follows:

$$V(x)=\sum_{j=1}^{N}U_j\delta(x-x_j) \qquad (1)$$

Thus, $U_j$ and $x_j$ represent the strength and the position of the $j$th delta function respectively, and N is the number of the Dirac delta functions in Dirac Comb. The distance between the adjacent barriers are given by $d_j=x_{j+1}-x_j$. The Schrödinger wave equation of one dimension can be written as follows:

$$-\frac{\hbar^2}{2m^*}\frac{d^2\psi(x)}{dx^2}+V(x)\psi(x)=E\psi(x) \qquad (2)$$

Thus, $V(x)$ is the periodic potential given by equation (1), $m^*$ is the electron effective mass, which is considered approximately constant over the interaction range. The solution of Schrödinger wave equation for single Delta function potential can be found in the literature and also in the transfer matrix formulism [14-16, 17]. The transfer matrix for periodic structure has been used also to study the transmission of electron through Comb structure [4-6, 14-16, 17]. The transfer matrix, which is related to the input electron wave and the output, is given by [4-6]

$$\begin{pmatrix}C\\D\end{pmatrix}=\begin{pmatrix}e^{ikx_j}&0\\0&e^{-ikx_j}\end{pmatrix}^{-1}\begin{pmatrix}1-i\beta_j&-i\beta_j\\+i\beta_j&1+i\beta_j\end{pmatrix}\begin{pmatrix}e^{ikx_j}&0\\0&e^{-ikx_j}\end{pmatrix}\begin{pmatrix}A\\B\end{pmatrix} \qquad (3)$$

Thus, $\beta_j$ is $\gamma_j/2k$, where $\gamma_j$ is $2m^*U_j/\hbar^2$, and $k$ is the wave number given by $\sqrt{2m^*E/\hbar^2}$. So that the transfer matrix at a given single barrier can be written as,

$$M_j=S^{-1}(ikx_j)\cdot\Gamma(\beta_j)\cdot S(ikx_j) \qquad (4)$$

The total transfer matrix which represents the electron propagating through the entire device is just the repetitive product of the transfer matrix of a single barrier. We find

$$M_t=S^{-1}(ikd_N)\cdot\Gamma(\beta_N)\cdot\left[\prod_{j=N-1}^{1}S(ikd_j)\cdot\Gamma(\beta_j)\right]\cdot S(0) \qquad (5)$$

Thus, $d_j$ is the periodic spacing between two adjacent Dirac Delta functions.
Then the transmission amplitude is given by [18],

$$T=\frac{1}{M_t(2,2)} \qquad (6)$$

Thus, $M_t(2,2)$ is the second element in the second row in a $2\times 2$ matrix. According to the Landauer-Buttiker formula, the electron conductance through this structure is [19,20]

$$G=\frac{2e^2}{h}|T|^2 \qquad (7)$$

We assume a dimensionless strength for Delta function potential [21] by rescaling our parameters, $\Omega_j=md_jU_j/\pi^2\hbar^2$. In figure (1), we show the conductance through $N=5$ Dirac delta function potential with strength $\Omega=0.2$. A perfect transmission in this case is in general impossible as predicted by [4, 22]. According to reference [21] we can not have a resonant transmission, $T=1$, even if N is very large.

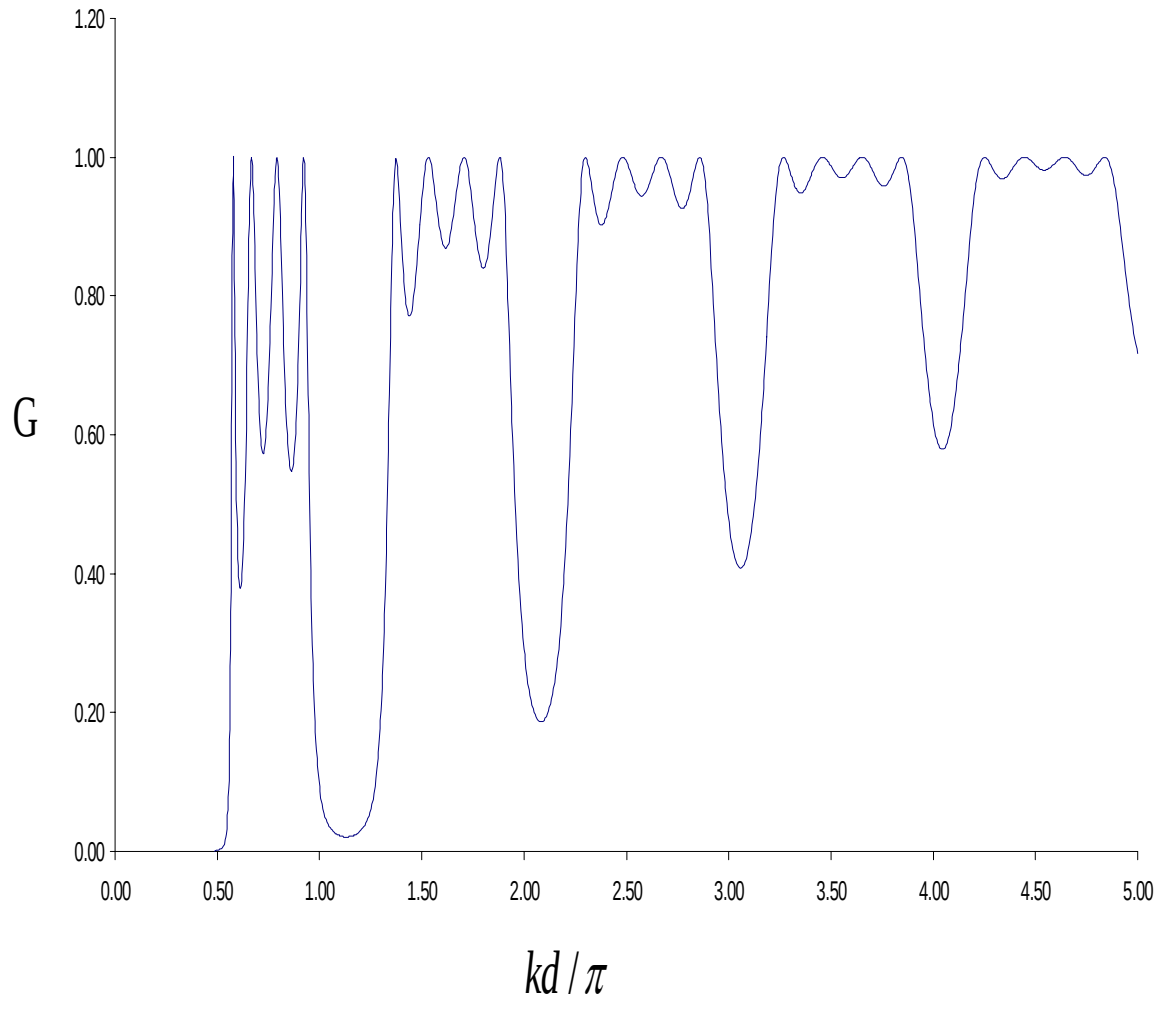

Figure 1: Conductance spectrum G in the units of $2e^2/h$ as a function of $kd/\pi$ for a sequence of Dirac delta function potential with N=5. The strength of the potential here is $\Omega=0.2$.

### 2.2 The Binomially Tailored Quantum Wire (BTQW)

In this subsection, we reintroduce BTQW structure as shown in figure 2. The Dirac delta function has been equally spaced but their strength, $\Omega_j$, has been weighted according to the binomial distribution law, which is

$$\Omega(N_j)=\binom{N}{N_j}/2^N,\ N_j=0,1,....,N \qquad (8)$$

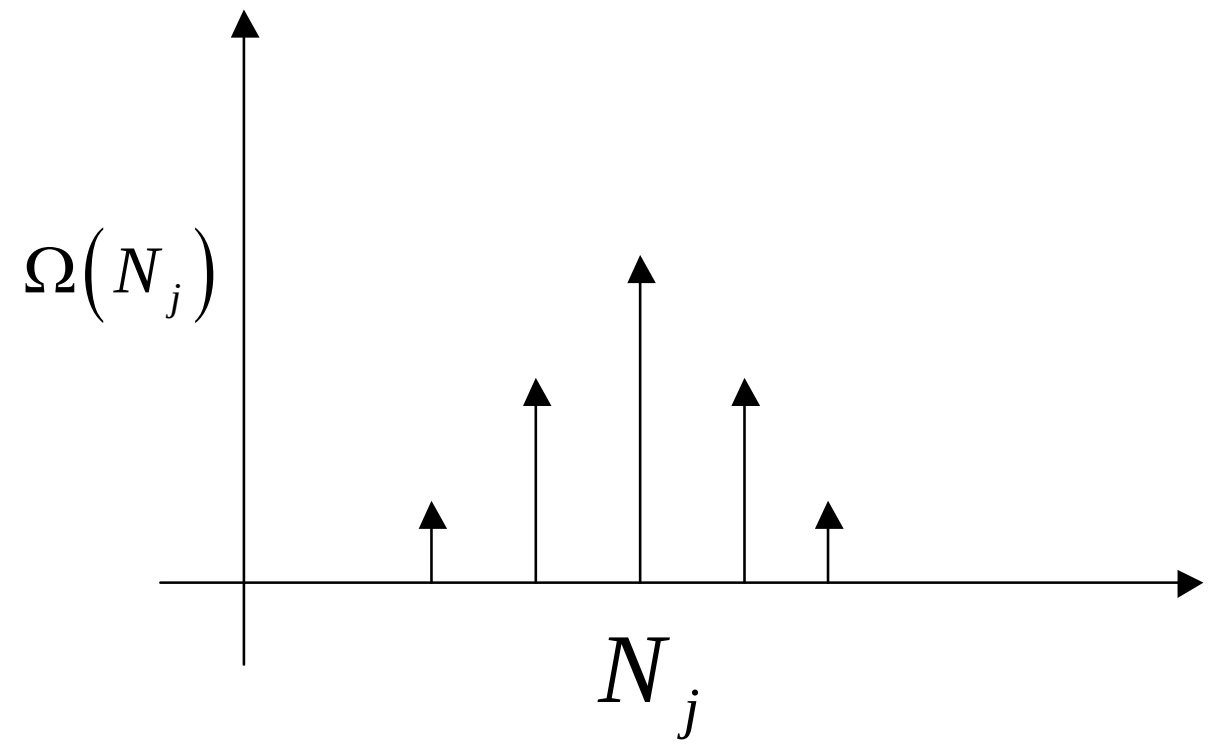

Figure 2: Shows binomially tailored Dirac delta function potential. $N_j$ values weighted by equation (8).

Thus, $\Omega\left(N_j\right)$ represents the strength of the Dirac delta potential, $N+1$ represents the total number of Dirac delta function potentials in the quantum wire, and $N_j$ is the order of the Dirac delta potential. This novel structure of quantum wires can be released by putting metallic gates on top of a one dimensional electron gas and then by applying voltages, according to the binomial distribution law, to deplete the electron gas underneath. In this case, equation (9) is no longer valid for our new structure. So that the total transmission matrix can be written as follows:

$$M_t = S^{-1}(ikd_{N+1})\cdot\Gamma(\beta_{N+1})\cdot S(ikd_N)\cdot\Gamma(\beta_N)\times \\ \text{..........}\times S(ikd_1)\cdot\Gamma(\beta_1)\cdot S(0) \qquad (9)$$

Notice that the potential strength is weighted according to equation (8). In figure 3, we show the conductance spectrum through a sequence of a binomially tailored Dirac delta function potentials. It is quite interesting to notice that we have reached a transmission through this structure approaches to unity in the allowed band region without any ripples after some $k$ value. Here, we have a resonant tunneling due to coherent interference effects due to elastic scattering of electrons, which leads the transmission to reach unity and also to have constant value over the allowed band or conduction band. Also, we see that there is a forbidden band or conduction gap where the transmission is small.

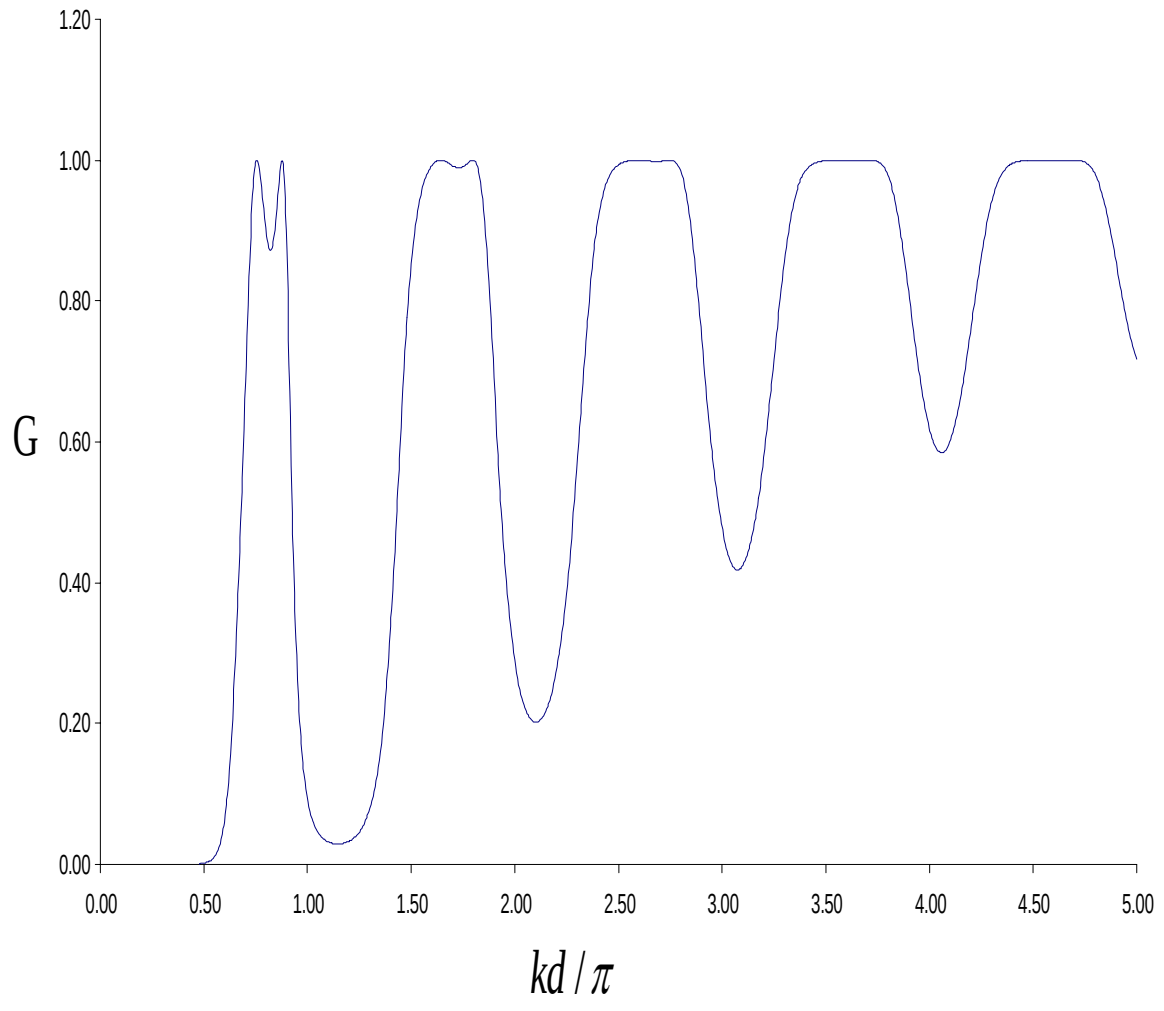

Figure 3: Conductance spectrum G in the units of $2e^2/h$ as a function of $kd/\pi$ for a BTQW, with N=5.

## 3. DEFECTS EFFECT ON THE ELECTRONIC CONDUCTANCE OF BTQW

### 3.1 Strength Defect

In this subsection, we study effect of strength defect of the central element of the binomial tailored quantum wire, and keeping the other elements and the spacing between the Dirac delta function potentials constant, on the electronic conductance through the BTQW. First, we consider the strength defect does not exceed $\pm 5\%$ of the Dirac delta function potential strength. That is, when the central Dirac delta functions potentials strength is $\Omega_j\left(N/2+1\right) \pm 0.05\Omega_j\left(N/2+1\right)$. In figure 4-a, we plot the electronic conductance spectrum for both strengths with $N_j$ *is* 35 and scaling factor of three. As can noticed there is slight difference between the two conductance spectrum curves, and defect free curves. In figure 4-b, we have increased the strength defect up to $\pm 20\%$, we have noticed some measurable differences between the two the conductance spectrum curves and defect free curves, but still the conduction band and the forbidden bands well defined, which is a very good feature for BTQW.

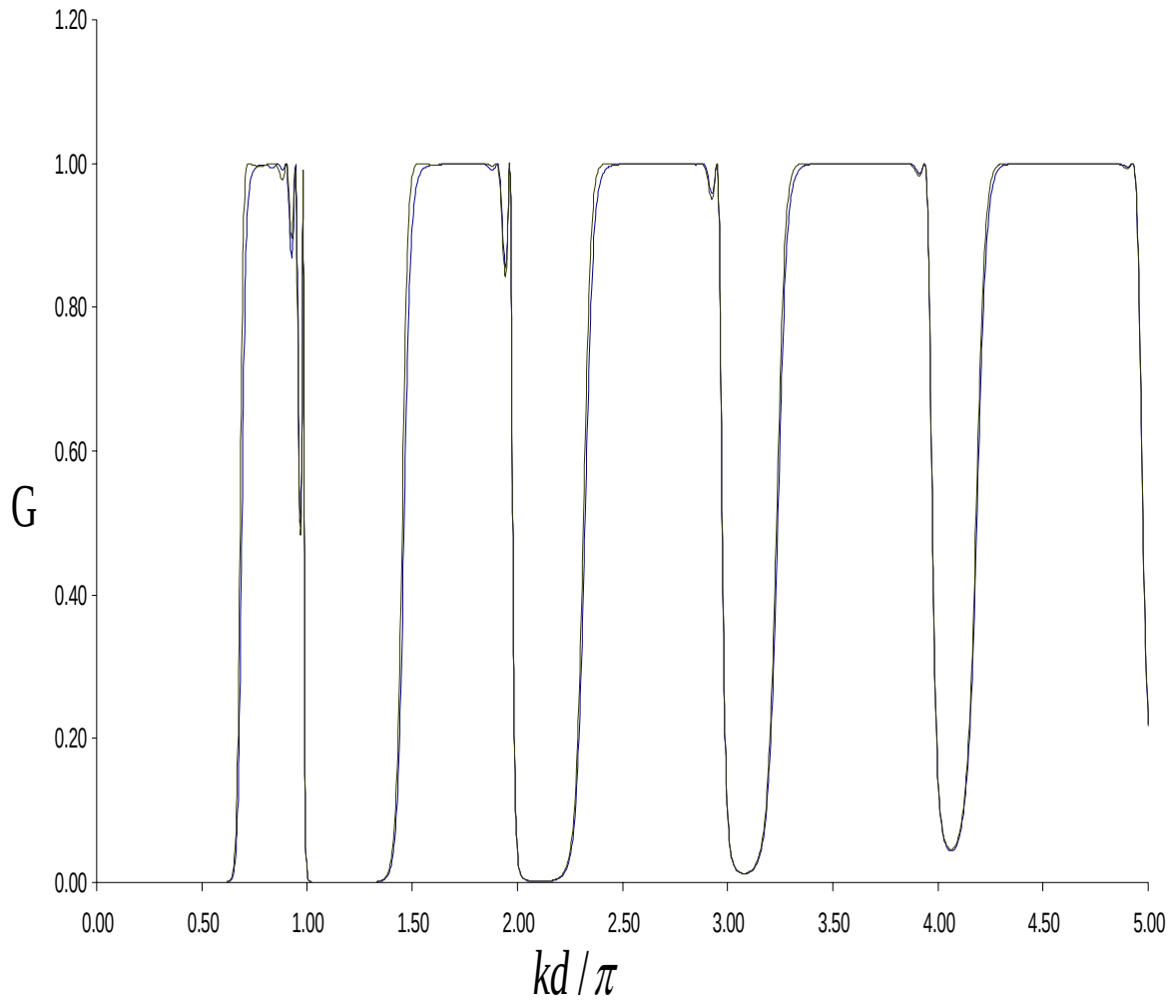

Figure 4-a: The electronic conductance, in the units of $2e^2/h$ as a function of $kd/\pi$. In this case, the defect is only $\pm 5\%$, in the strength of the central Dirac delta function.

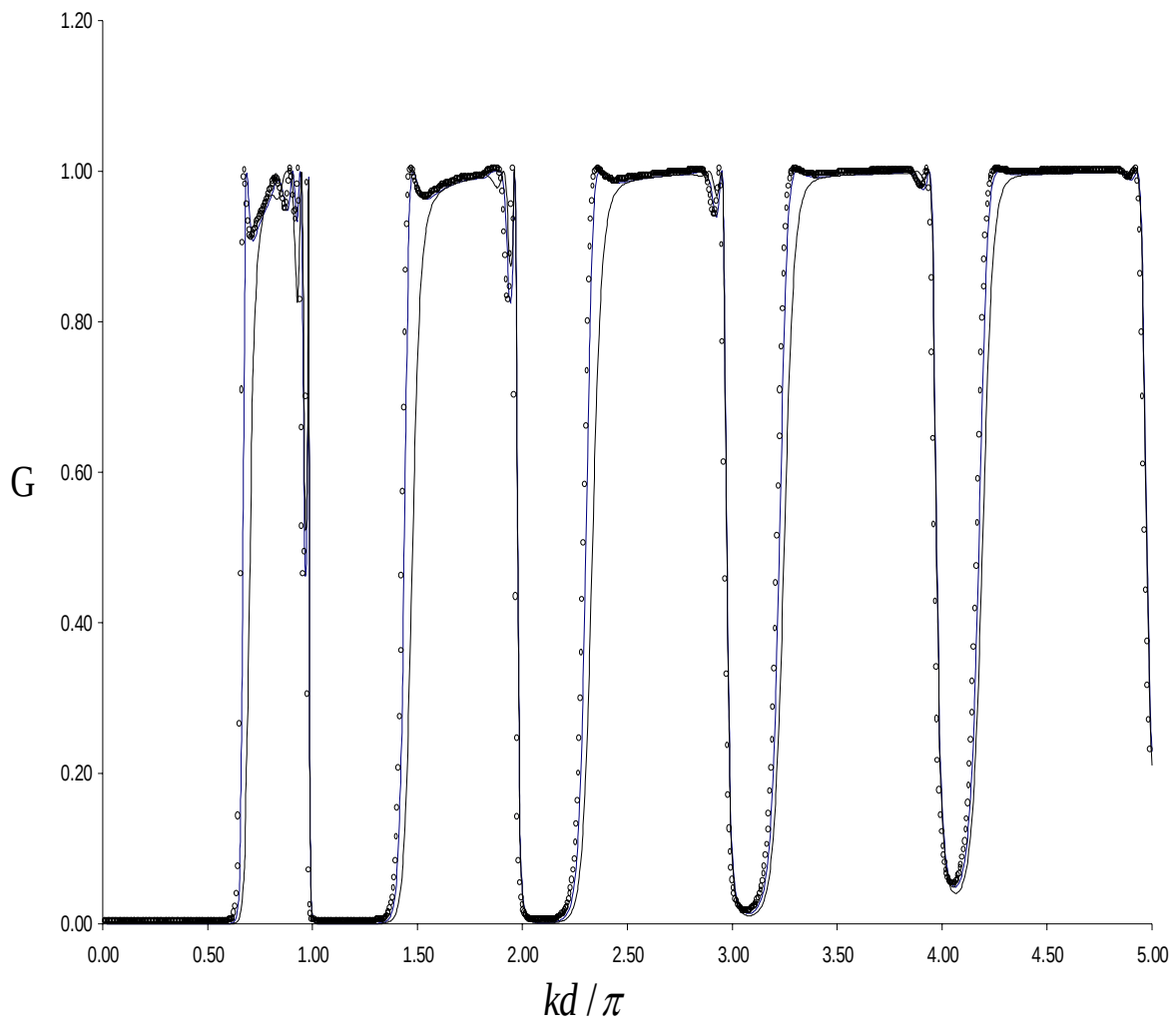

Figure 4-b: The electronic conductance, in the units of $2e^2/h$ as a function of $kd/\pi$. In this case, the defect is only $\pm 20\%$, in the strength of the central Dirac delta function.

## 3.2 Dislocation Effect

In this subsection, we study dislocation defect effect on the position of the central element in the BTQW, and keeping the other elements and the spacing between the Dirac delta function potentials constant. First, we consider the position defect does not exceed $\pm 5\%$ of the Dirac delta function potential spacing constant. That is, when the central Dirac delta function potentials spacing is $d \pm 0.05d$. In figure 5-a, we plot the electronic conductance spectrum for both dislocations with $N_j$ *is* 35 and scaling factor of three. Compared to defect curves, as can noticed there is a difference between the two curves. The conduction band starts lose its flatness and the forbidden band shaper for increased spacing between the central Dirac delta function and the adjacent one. In figure 5-b, we increase the dislocation defect up to $\pm 20\%$, we have noticed measurable differences between the two curves and that of no defect case, but still the conduction band is well defined but the forbidden bands have a split compared to forbidden band in no defect curves. This splitting is due to resonant state in the forbidden energy band which leads to a bound state in the structure [3]. This is because the particle mode cannot propagate and hence get trapped.

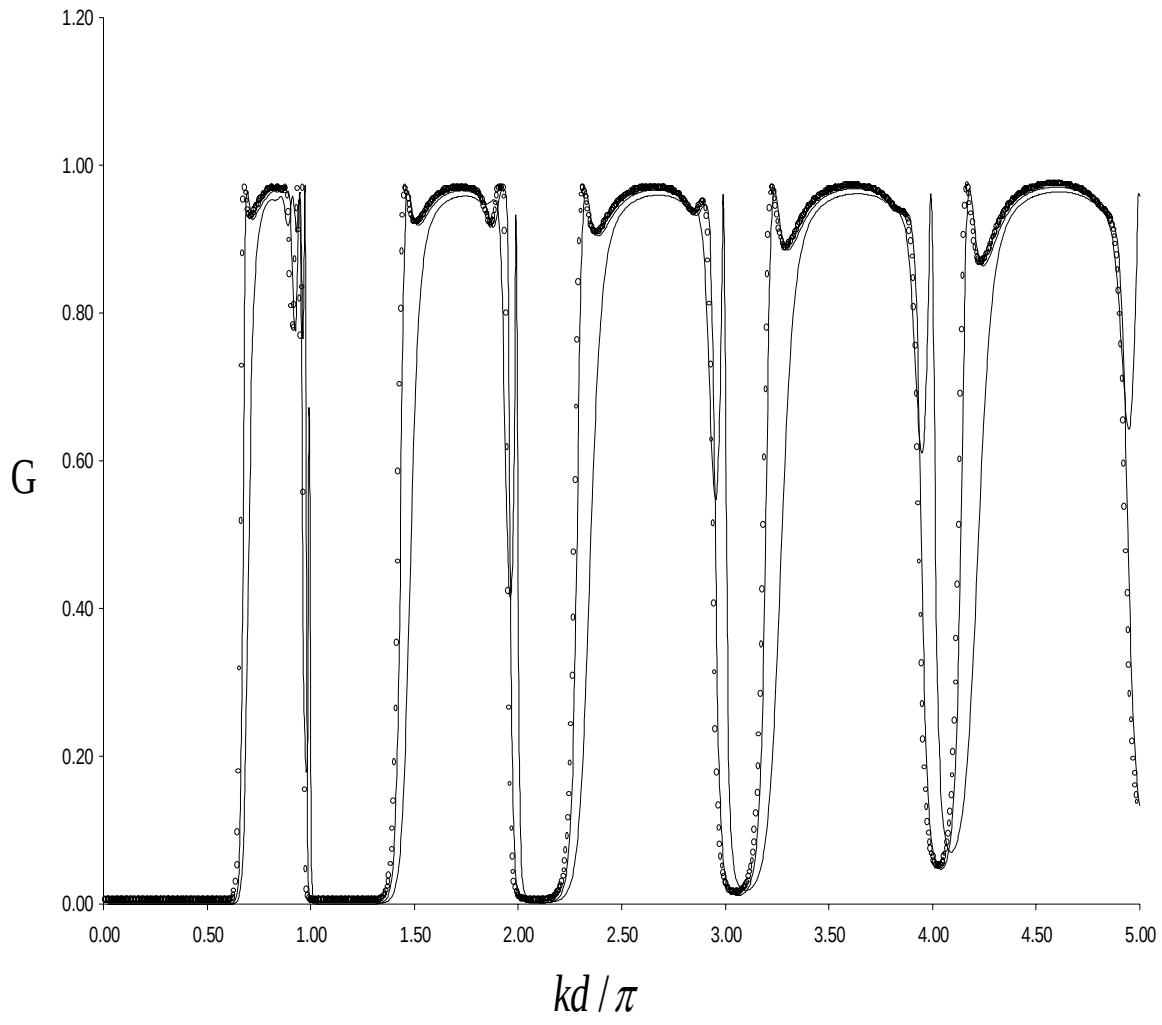

Figure5-a: The electronic conductance, in the units of $2e^2/h$ as a function of $kd/\pi$. In this case, the defect is only $\pm 5\%$, in the position of the central Dirac delta function.

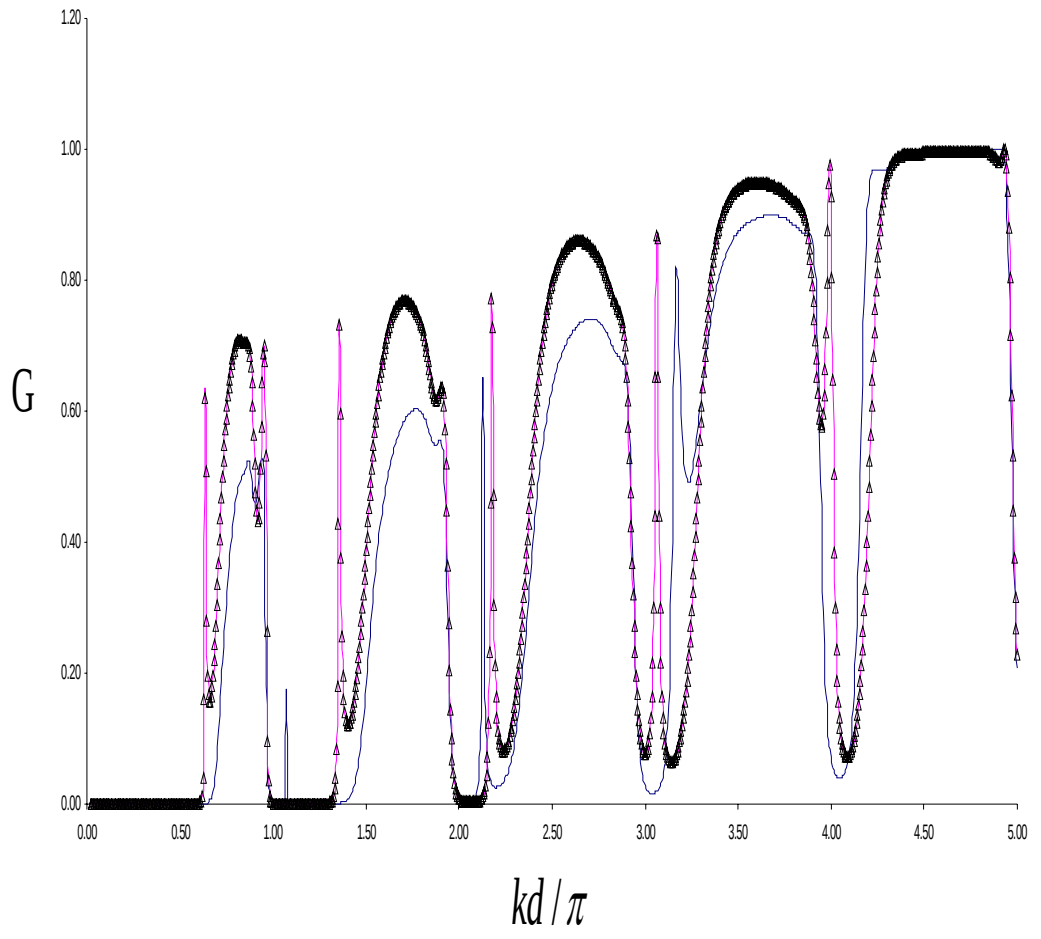

Figure 5-b: The electronic conductance, in the units of $2e^2/h$ as a function of $kd/\pi$. In this case, the defect is only $\pm 20\%$, in the position of the central Dirac delta function.

## 4. CONCLUSION

We found the novel structure introduced by [13] has a good tolerance for strength reaches up to $\pm 20\%$ and dislocation tolerance reaches up to $\pm 5\%$ without losing the fascinating electronic transmission characteristics.

## REFERENCES

[1] D. W. L. Sprung, and H. Wu, Am. J. Phys. **61** (12), 1118, December 1993.
[2] D. Kiang, Am. J. Phys. **42**, 785-7 (1974).
[3] P. Singha Deo, and A. M. Jayannavar, Phys. Rev. B 50,11 629 (1994).
[4] H. Fayad, M. M. Shabat, Islamic Journal of Gaza (Natural Sciences Series) vol.13, no.2, P.203-211, 2005
[5] H. Fayad, M. M. Shabat, H. Khalil, and D. Jäger, Proceeding of the IEEE Electron Devices Soc., IEEE, vol.11, p.91-4, 2001.
[6] H. Fayad, and M. M. Shabat, "Electronic Conductance through some quantum wire structure", Micro and Nano-Engineering Series, Nano science and Nano engineering, Romanian Academy, eds., Dan Dascalu and Irina Kleps, Bucharest, Romania, 2002
[7] D. J. Griffith and N. F. Taussing, Am. J. Phys. **60**, 883-8 (1992).
[8] D. J. Vezzetti and M. Cahay, J. Phys. D 19, L53-55 (1986).
[9] G. J. Jin, Z. D. Wang, A. Hu and S.S. Jiang, J. Appl. Phys., Vol. 85, No. 3, 1597 (1999)
[10] M. Macucci, A. Galick, and U. Ravailoli, Phys Rev. B 52, 5210 (1995).
[11] Y. Takagaki, and D.K. Ferry, Phys Rev. B 45, 6715 (1992).
[12] H. Tachibana and Totsuji, J. Appl. Phys. 79, 7021 (1996).
[13] H. S. Ashour, A.I. Assad, M. M. Shabat, and M. S. Hamada, Microelectronic Journal (in press)
[14] Hua Wu, D. W. L. Sprung, J .Martorell, and S. Klarsfeld, Phys. Rev B 44, 6351 (1991).
[15] W. D. Sheng, and J. B. Xia, J. Phys.:Condens Matter 8, 3635 (1996).
[16] T. Kostyrko, Phys. Rev. B 62, 2458 (1999).
[17] E. Merzbacher, Quantum Mechanics ( Wiley, New York, 1970).
[18] D. W. L. Sprung, Hua Wu, and J .Martorell, Am. J. Phys. 61, 1118 (1993).
[19] L. D. Landau and E. M. Lifshitz, Quantum Mechanics ( Pergamon Press, Oxford, 1976).
[20] G. Baym, Lectures on Quantum Mechanics (W. A. Benjamin, Inc., Massachusetts, 1973).
[21] Y. Takagaki, and D.K. Ferry, Phys Rev. B 45, 8506 (1992).
[22] S. J. Blundell, Am. J. Phys. 61, 1147 (1993).